\documentclass[onecolumn,10pt]{IEEEtran}

\usepackage{indentfirst,amsmath}
\usepackage{color,rotating,pdflscape}
\usepackage{amsfonts,amsmath,amssymb,amsthm}
\usepackage{graphicx,multirow,bm}
\usepackage{cite}
\usepackage{diagbox}
\usepackage{float}
\makeatletter
\newtheoremstyle{mythm}{3pt}{3pt}{}{16pt}{\bfseries}{:}{.5em}{}
\theoremstyle{mythm}
\newtheorem{theorem}{Theorem}
\newtheorem{example}{Example}

\newtheorem{corollary}{Corollary}
\newtheorem{lemma}{Lemma}

\begin{document}
\title{Reduced Transmission in Multi-Server Coded Caching
\author{Minquan Cheng, Qiaoling Zhang, Jing Jiang, Ruizhong Wei
\thanks{M. Cheng, Q. Zhang and J. Jiang are with Guangxi Key Lab of Multi-source Information Mining $\&$ Security, Guangxi Normal University,
Guilin 541004, China (e-mail: $\{$chengqinshi,qlzhang2017,jjiang2008$\}$@hotmail.com).}
\thanks{R. Wei is with Department of Computer Science, Lakehead University, Thunder Bay, ON, Canada, P7B 5E1.
             (e-mail: rwei@lakeheadu.ca) }
}}
\date{}
\maketitle

\begin{abstract}
Coded caching has been widely used in the wireless network for shifting the some transmissions during the peak traffic times to the off-peak traffic times. Multi-server coded caching, which can share responsibility for the total amount of transmission in the wireless network during the peak traffic times by means of the collaboration among these servers, can be seen everywhere in our life. The three servers setting (two data servers and one parity check server) is  used in practice, e.g. redundant array of independent disks-4. In this scenario, there are total $N$ files which are equally stored in two data servers respectively and $K$ users each of which has the memory size of $M$ files. Each server connects to users by an independently channel. During the off-peak traffic times, two data servers place some parts of each files in each user's cache. In that time, servers do not know users' requests in future. During the peak traffic times each user just requests one file from $N$ files.
Luo et al. in 2016 proposed the first coded caching scheme for this setting. In this paper, we proposed some method that further reduces the amount of transmission in each channel when $\frac{KM}{N}$ is odd. This method also improves the transmission rate for systems with general multiply servers.
\end{abstract}

\section{Introduction}
\label{intro}
Predominantly driven by video content demand, there is a dramatic increase in wireless traffic now. The high temporal variability of network traffic results in communication systems that are congested
during peak-traffic times and underutilized during off-peak times. Caching is a natural strategy to cope with this high temporal variability by shifting some transmissions from peak to off-peak times with the help of cache memories
at the network edge.

Maddah-Ali and Niesen in \cite{MN} proved that coded caching does not only shift some transmissions from peak to off-peak times, but also further reduces the amount of transmission during the peak traffic times by exploiting caches to create multicast opportunities. The first caching scenario  focused in \cite{MN} is: a single server containing $N$ files with the same length connects to $K$ users over a shared link and each user has a cache memory of size $M$ files. During the off-peak traffic times the server places some contents to each user's cache.
In this phase the server does not known what each user will require next. During the peak traffic times, each user requires a file from server randomly. Then according to each user's cache, the server sends a coded signal (XOR of some required packets) to the users such that various user demands are satisfied.
The first determined coded caching scheme, which is called MN scheme in this paper, was proposed in \cite{MN}. It is worth mentioning that the broadcasted amount of MN scheme for the worst request, where all the requirements are different from each other, is at most four times larger than the lower bound when $K\leq N$ \cite{GR}. We denote such amount by $R_{MN}(K,\frac{M}{N})$. So MN scheme has been extensively employed in practical scenarios, such as device to device networks \cite{JCM}, hierarchical networks \cite{KNMD}, security \cite{STC}, multi-servers setting \cite{LAP,SMK,MGL} and so on.  There are also many results following MN scheme in \cite{GR,T,WLG,WTP1,WTP,YCTC,YMA} etc.

The coded caching used in muti-server setting can be seen everywhere. We focus on the setting in \cite{LAP} which is also widely used (e.g. redundant array of independent disks-4) in our life. In this setting there are three servers, i.e., two data servers $A$, $B$ storing $N/2$ disjoint files respectively and a parity server $P$ storing the bitwise XOR of the information in $A$ and $B$. The servers connect to users and operate on independent errorfree channels. This implies that these servers can transmit messages simultaneously and without interference to the same or different users. In practice servers are aware of the content cached by each user and of the content stored in other servers.
So even though any two files sorted on different servers can not be combined into a single message, the servers can still coordinate the messages of these two files. Similar to the single server setting, assume that each user request one file from $N$ files and sends to the three servers. Then each server combines multiple segments from its own files into a single message, and broadcasts them respectively such that each user can be satisfied by means of its cache and the received signal messages from servers. We prefer that the amount of transmission in each channel is as small as possible. Denote the maximum amount broadcasted among the three servers by $R$ files for all the requests. Clearly it is meaningful to design a scheme such that $R$ is as small as possible. $R$ is referred to the rate of a scheme.

Luo et al., in \cite{LAP} constructed the first determined coded caching scheme by means of MN scheme and the results on saturating matching in bipartite graph. Specifically they first considered the symmetric request, i.e., both data servers receive the same number of requests, and showed that in their scheme the rate $R=\frac{1}{2}R_{MN}(K,\frac{M}{N})$ if $\frac{KM}{N}$ is even, otherwise $R=(\frac{1}{2}+\frac{1}{6})\Delta R_{MN}(K,\frac{M}{N})$ where the upper bound of $\Delta$ is $\frac{1}{3}$. Then a scheme and the related rate for the other requests  can be obtained directly by means of several classes of schemes in symmetric requests.

\begin{figure}
  \centering
  \includegraphics[width=0.8\textwidth]{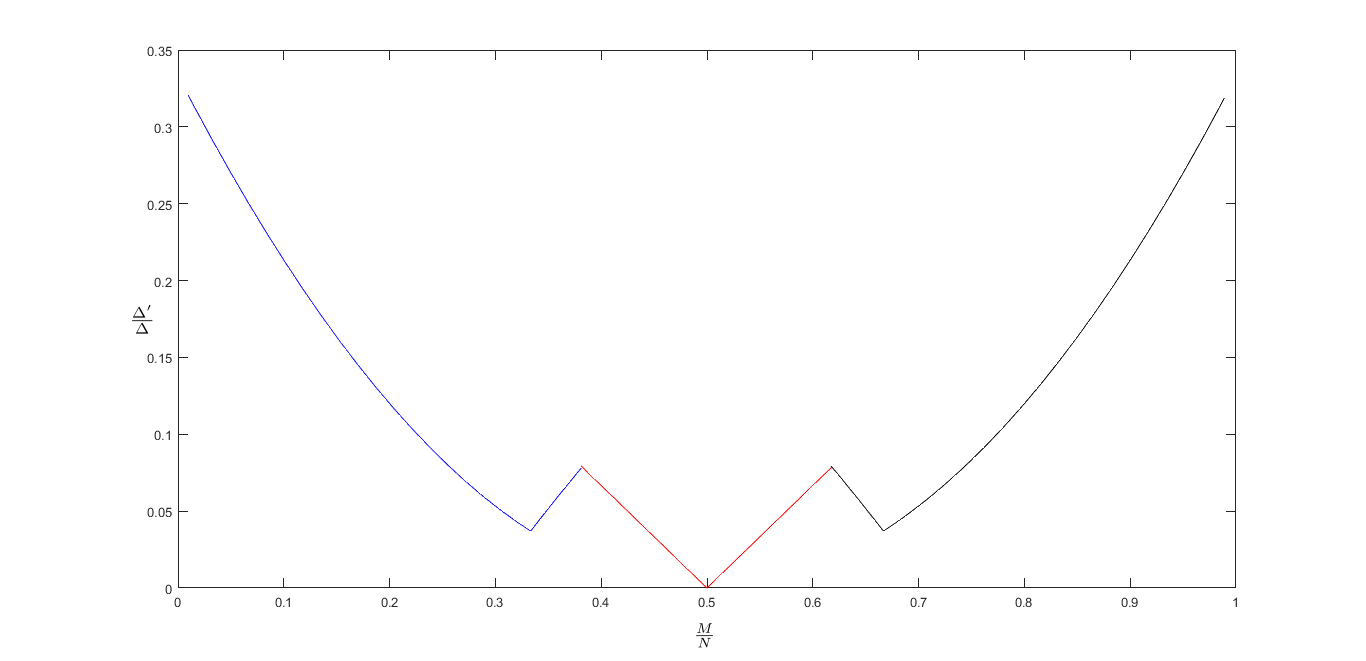}
  \caption{The function $\frac{\Delta'}{\Delta}$ when $\frac{KM}{N}$ is an odd integer\label{fig-all}}
\end{figure}

In this paper, we further investigate the caching system with three servers, i.e., there are two data servers, one parity check server connecting to $K$ users independently. By modifying the schemes in \cite{LAP} we derive a new rate $R=(\frac{1}{2}+\frac{1}{6}\Delta' )R_{MN}(K,\frac{M}{N})$ when $\frac{KM}{N}$ is an odd integer.  Figure \ref{fig-all} indicated that $\Delta' $ is obviously smaller than $\Delta$ in most cases.
In particular when  $K$ is large, $\frac{\Delta'}{\Delta}\approx 0$ if $\frac{M}{N}=\frac{1}{2}$, and $\frac{\Delta'}{\Delta}\approx\frac{1}{9}$ if $\frac{M}{N}=\frac{1}{3}$ or $\frac{2}{3}$.
The rest of this paper is organized as follows. Section \ref{preliminaries} briefly reviews MN scheme, the scheme proposed in \cite{LAP} and the related concepts. In Section \ref{sec-Sepcial-idea}, an improved scheme and its performance analysis are proposed for the case of
three servers. Then
we discussed briefly the general cases in Section IV. Conclusion is drawn in Section \ref{conclusion}.

\section{Preliminaries}\label{preliminaries}
\subsection{MN scheme and bipartite graph}
We consider a network of $K$ users and $N$ files,  denote by $W_{1}$, $W_{2}$, $\ldots, W_N$, such that each user has a cache with capacity for $M$
files. We denote that network as
 a $(K,M,N)$ caching system.  In the single server setting, when $t=\frac{KM}{N}$ is an integer, an MN scheme can be described as follows \cite{MN}.
\begin{itemize}
\item During the off-peak traffic times,  file $W_i$ is divided into $F={K\choose t}$ equal packets, so that  $W_{i}=\{W_{i,\mathcal{T}}\ |\ \mathcal{T}\subseteq [1,K],|\mathcal{T}|=t\}$. User $k$ caches the following packets.
\begin{eqnarray*}
\mathcal{Z}_k=\{W_{i,\mathcal{T}}\ |\ k\in \mathcal{T}, i=1,2,\ldots,N\}
\end{eqnarray*}
\item During the peak traffic times, each user requires a file randomly. Suppose  the users request  ${\bf d}=(d_1,d_2,\ldots,d_K)$. Then for each
$t+1$ subset  of users $\mathcal{S}$, the server sends the following coded signal to each user of $\mathcal{S}$ .
\begin{eqnarray}
\label{eq-MN-De}
\bigoplus_{k\in \mathcal{S}}W_{d_k,\mathcal{S}\setminus\{k\}}
\end{eqnarray}

Clearly the server broadcasts ${K\choose t+1}$ times. So the amount of transmission by server is
\begin{eqnarray*}
\label{eq-MN-rate}
R_{MN}(K,{\textstyle\frac{M}{N}})={K\choose t+1}\Big/{K\choose t}=\frac{K-t}{t+1}.
\end{eqnarray*}
\item
Each user  uses the cached segments to recover the designed segment and then the requested file.
\end{itemize}

 A graph is denoted by $\mathbf{G}=(\mathcal{V},\mathcal{E})$, where $\mathcal{V}$ is the set of vertices and $\mathcal{E}$ is the set of edges.
A subset of edges $\mathcal{M}\subseteq \mathcal{E}$ is a matching if no two edges have a common vertex. A bipartite graph, denoted by $\mathbf{G}=(\mathcal{X},\mathcal{Y}; \mathcal{E})$, is a graph whose vertices are divided into
two disjoint parts $\mathcal{X}$ and $\mathcal{Y}$ such that every edge in $\mathcal{E}$ connects a vertex in $\mathcal{X}$ to one in $\mathcal{Y}$. For a set $X\subseteq \mathcal{X}$, let $N_{\mathbf{G}}(X)$ denote the set of all vertices in $\mathcal{Y}$ adjacent to some vertex of $X$.
The degree of a vertex is the number of vertices adjacent to it. If every vertex of $\mathcal{X}$ has the same degree, we also call such a degree the degree of $\mathcal{X}$ and denote $d(\mathcal{X})$.
\begin{theorem}\rm(Hall's Marriage Theorem, \cite{BM} )
\label{th-hall}
Given a bipartite graph $\mathbf{G}=(\mathcal{X},\mathcal{Y};\mathcal{E})$, there exists a matching with $|\mathcal{X}|$ edges, i.e., a saturating matching, if and only if $|X|\leq |N_{\mathbf{G}}(X)|$ holds for any subset $X\subseteq\mathcal{X}$.
\end{theorem}
\begin{corollary}\rm(\cite{BM})
\label{co-Hall}
Given a bipartite graph $\mathbf{G}=(\mathcal{X},\mathcal{Y};\mathcal{E})$, assume that $d(\mathcal{X})=m$ and $d(\mathcal{Y})=n$. If $m\leq n$, then there is a saturating matching.
\end{corollary}
\subsection{The scheme in \cite{LAP}}\label{LAP}

 \cite{LAP} proposed a simple caching system with 3 severs, two data servers and a third server storing their bitwise XOR.
In the following we denote the files in server $A$ and $B$ by $\{A_{1}$, $\ldots$, $A_{N/2}\}$ and $\{B_{1}$,$\ldots$, $B_{N/2}\}$ respectively.
So the files in parity server $P$ are $\{A_{1}\oplus B_{1}$, $\ldots$, $A_{N/2}\oplus B_{N/2}\}$,  as showed in Table \ref{tab-server-system}.
\begin{table}
  \centering
  \caption{Files stored in each server} \label{tab-server-system}
  \normalsize{
\begin{tabular}{|c|c|c|}
\hline
Server $A$ & Server $B$ & Server $P$ \\ \hline
$A_1$      & $B_1$      & $A_1\oplus B_1$\\
$A_2$      & $B_2$      & $A_2\oplus B_2$\\
$\vdots$   & $\vdots$   & $\vdots$       \\
$A_{N/2}$  & $B_{N/2}$  & $A_{N/2}\oplus B_{N/2}$\\ \hline
\end{tabular}}
\end{table}
\noindent
The following notations will be used.
\begin{itemize}
\item Denote the set of users requesting files from server $A$ by $\mathcal{K}_A$, and the set of users requesting files from server $B$ by $\mathcal{K}_B$. Clearly the set of all the users is $\mathcal{K} = \mathcal{K}_A \bigcup \mathcal{K}_B$. Let $K_A=|\mathcal{K}_A|$ and $K_B=|\mathcal{K}_B|$.
\item Assume that the $k$th user of $\mathcal{K}_A$ requests the $d_{k}$-th file in server $A$, and the $k$th user of $\mathcal{K}_B$ requests the
$d_{k}$-th file in server $B$.
\end{itemize}
Following \cite{LAP}, we only consider the case that $t=\frac{KM}{N}$ is an integer for the simplicity. Luo et al., in \cite{LAP} used the same caching strategy as MN scheme during the off-peak traffic times for each server,  but modified the coded signals in \eqref{eq-MN-De} during the peak traffic times as follows. Given a subset of users $\mathcal{S}_1$ of size $t+1$, it can be divided into three parts, say $\mathcal{Q}_A$, $\mathcal{Q}_B$ and $\mathcal{Q}_{A}^{'}$ where $\mathcal{Q}_A,\mathcal{Q}_{A}^{'}\subseteq \mathcal{K}_A$ and  $\mathcal{Q}_B\subseteq \mathcal{K}_B$.
If there exists another subset $\mathcal{S}_2$ of size $t+1$ which can be divided into $\mathcal{Q}_A$, $\mathcal{Q}_B$ and $\mathcal{Q}_{B}^{'}$
where $\mathcal{Q}_{B}^{'}\subseteq \mathcal{K}_B$,
then the pair $(\mathcal{S}_1,\mathcal{S}_2)$ is called an effective pair.
When servers $A$, $B$ and $P$ broadcast the following messages respectively
\begin{small}
\begin{eqnarray}
\label{eq-broa-mthod}
\begin{split}
 m_{\mathcal{S}_1}^A&={\textstyle\left(\bigoplus\limits_{k\in \mathcal{Q}_A}A_{d_k,\mathcal{S}_1\setminus \{k\}}\right)}{\textstyle \bigoplus}
\left({\textstyle\bigoplus\limits_{k\in \mathcal{Q}_B}A_{d_k,\mathcal{S}_1\setminus \{k\}}}\right){\textstyle\bigoplus}
\left({\textstyle\bigoplus\limits_{k\in \mathcal{Q}_A'}A_{d_k,\mathcal{S}_1\setminus \{k\}}}\right)
\\[0.2cm]
m_{\mathcal{S}_2}^B&={\textstyle\left(\bigoplus\limits_{k\in \mathcal{Q}_A}B_{d_k,\mathcal{S}_2\setminus \{k\}}\right)\bigoplus
\left(\bigoplus\limits_{k\in \mathcal{Q}_B}B_{d_k,\mathcal{S}_2\setminus \{k\}}\right)\bigoplus
\left(\bigoplus\limits_{k\in \mathcal{Q}_B'}B_{d_k,\mathcal{S}_2\setminus \{k\}}\right)}\\[0.2cm]
m_{\mathcal{S}_1\bigcap \mathcal{S}_2}^P&={\textstyle
\left[\bigoplus\limits_{k\in \mathcal{Q}_B}\left(A_{d_k,\mathcal{S}_1\setminus \{k\}}\oplus B_{d_k,\mathcal{S}_1\setminus \{k\}}\right)\right]\bigoplus
\left[\bigoplus\limits_{k\in \mathcal{Q}_A}\left(B_{d_k,\mathcal{S}_2\setminus \{k\}}\oplus A_{d_k,\mathcal{S}_2\setminus \{k\}}\right)\right]}\\
&={\textstyle\left[\left(\bigoplus\limits_{k\in \mathcal{Q}_B}A_{d_k,\mathcal{S}_1\setminus \{k\}}\right)\bigoplus
\left(\bigoplus\limits_{k\in \mathcal{Q}_B}B_{d_k,\mathcal{S}_1\setminus \{k\}}\right)\right]}\\
&\ \ \ \ {\textstyle\bigoplus
\left[\left(\bigoplus\limits_{k\in \mathcal{Q}_A}B_{d_k,\mathcal{S}_2\setminus \{k\}}\right)\bigoplus
\left(\bigoplus\limits_{k\in \mathcal{Q}_A}A_{d_k,\mathcal{S}_2\setminus \{k\}}\right)\right]}
\end{split}\end{eqnarray}
\end{small}
Then each user in $\mathcal{S}_1$ and $\mathcal{S}_2$ can obtain the requested segments from $m_{\mathcal{S}_1}^A$, $m_{\mathcal{S}_2}^B$ and $m_{\mathcal{S}_1\bigcap \mathcal{S}_2}^P$.
So if the sets $\mathcal{S}_1$ and $\mathcal{S}_2$ form an effective pair, then the messages indexed by $\mathcal{S}_1$ and $\mathcal{S}_2$ in \eqref{eq-MN-De} can be replaced by three messages in \eqref{eq-broa-mthod}.
 In the case of symmetric request, i.e., $K_A=K_B$, Luo et al., obtained the following results.
\begin{theorem}\rm(\cite{LAP} )
\label{th-rate-G}
Based on MN scheme, when $K_A=K_B$ and $K=K_A+K_B$, the rate of the server system in Table \ref{tab-server-system} is
\begin{eqnarray}
R_{T}(K,{\textstyle\frac{M}{N}})=\left\{\begin{array}{cc}
            \frac{1}{2}R_{MN}(K,\frac{M}{N}) & \hbox{if}\ \frac{KM}{N} \ \hbox{is even}\\
            (\frac{1}{2}+\frac{1}{6}\Delta ) R_{MN}(K,\frac{M}{N}) & \hbox{if}\ \frac{KM}{N} \ \hbox{is odd}
 \end{array}\right.
\end{eqnarray}
where $\Delta $, which is bounded by $\frac{1}{3}$, represents the ratio of unpaired messages.
\end{theorem}
In the following we will focus on the symmetric request.

\cite{LAP} investigated  the value of $\Delta$ (we will see it later) using their method. The main purpose of this paper is to give
some refined method that can improve (reduce) the value of $\Delta$. Using our method, the value of $\Delta $ depends on
$\lambda = M/N$. Therefore in what follows, we will discuss the value of  $\Delta$ according to different
 values of $\lambda$.

 To use graphs to discuss the values of $\Delta$, we will use the following settings.
Each vertex of a graph is always represented  a subset $\mathcal{S}\subseteq \mathcal{K}$ with size $t+1$. And for any bipartite graph $\mathbf{G}=(\mathcal{X}, \mathcal{Y};\mathcal{E})$ where a vertex $\mathcal{S}\in \mathcal{X}$ is adjacent to $\mathcal{S}'\in \mathcal{Y}$ if and only if they can form an effective pair.

\subsection{Research motivation}
A brief review of the  proof of Theorem \ref{th-rate-G} is  useful to understand our proofs.
Here we take the case that  $t=\frac{KM}{N}$ is odd as an example. For each $w=0$, $1$, $\ldots$, $t+1$, define 
\begin{eqnarray}
\label{eq-bi-XY}
\begin{split}
\mathcal{V}_w&=\{\mathcal{S}\subseteq \mathcal{K}\ |\ |\mathcal{S}|=t+1, |\mathcal{S}\bigcap\mathcal{K}_A|=w\}\\
\end{split}
\end{eqnarray}
Clearly $|\mathcal{V}_w|=|\mathcal{V}_{t+1-w}|={K_A\choose w}{K_B\choose t+1-w}={K_A\choose t+1-w}{K_B\choose w}$ since $K_A=K_B$. By the fact ${K\choose t+1}=\sum_{w=0}^{t+1}{K_A\choose w}{K_B\choose t+1-w}$,
Luo et. al. \cite{LAP}  constructed several classes of bipartite graphes satisfying Corollary \ref{co-Hall} in the following way. For each $w\in[1,\frac{t-1}{2})$, they defined a bipartite graph
$\mathbf{G}_w=(\mathcal{V}_w,\mathcal{V}_{t+1-w};\mathcal{E}_w)$ by \eqref{eq-bi-XY} and
showed that these bipartite graphes satisfy Corollary \ref{co-Hall}. When $K_A<t+1$, $\mathcal{S}\bigcap \mathcal{K}_A\neq \emptyset$ always holds for each subset $\mathcal{S}\subseteq \mathcal{K}$ with cardinality $t+1$. So they did not need to consider the case $w=0$. When $K_A\geq t+1$ and $w=0$, assume that server $A$ (and $B$) broadcast the messages $m^A_{\mathcal{S}}$ (and $m^B_{\mathcal{S}}$), $\mathcal{S}\subseteq \mathcal{K}_A$ (and $\mathcal{S}\subseteq \mathcal{K}_B$) independently. For the sets $\mathcal{V}_{\frac{t-1}{2}}$, $\mathcal{V}_{\frac{t+1}{2}}$, and $\mathcal{V}_{\frac{t+3}{2}}$, let
\begin{eqnarray}
\label{eq-bi-XY-L}
\begin{split}
\mathcal{X}=\mathcal{V}_{\frac{t-1}{2}}\bigcup \mathcal{V}_{\frac{t+3}{2}},\ \ \ \ \ \ \
\mathcal{Y}=\mathcal{V}_{\frac{t+1}{2}}.
\end{split}
\end{eqnarray}
They defined a bipartite graph $\mathbf{G}=(\mathcal{X},\mathcal{Y};\mathcal{E})$ and showed that there is a saturating matching. So the number of unpaired messages is
\begin{small}
\begin{eqnarray*}
\label{eq-unpaired}
n=\left|{K_A\choose (t+1)/2}{K_B\choose (t+1)/2}- {K_A\choose (t-1)/2}{K_B\choose (t+3)/2}-{K_B\choose (t-1)/2}{K_A\choose (t+3)/2}  \right|
\end{eqnarray*}
\end{small}and the ratio of unpaired messages is
\begin{eqnarray}
\label{eq-L-Delta}
{ \Delta =\frac{n}{{K\choose t+1}}.}
\end{eqnarray}
Since each unpaired message can be transmitted by any two servers, each server could transmit $\frac{2}{3}n$ unpaired messages. So the rate is
{ \begin{eqnarray}
\label{eq-R-L-M}
R_{T}(K,{\textstyle\frac{M}{N}})=\frac{\left((1-\Delta+\frac{2}{3}\Delta\right){K\choose t+1}}{{K\choose t}}=\left(\frac{1}{2}+\frac{1}{6}\Delta\right)R_{MN}(K,{\textstyle\frac{M}{N}}).
\end{eqnarray}}
This is the result in Theorem \ref{th-rate-G}.
Next we will provide some method that can improve the ratio of unpaired messages so that to reduce the rate $R_{T}(K,\frac{M}{N})$.



\section{Improved scheme for three servers}
\label{sec-Sepcial-idea}
In this section, we focus on the case of symmetric request for the case that $t$ is odd. Clearly an intuitive approach to reduce the ratio of unpaired messages is finding the maximal matching of graph $\mathbf{G}=(\mathcal{V}_{\frac{t-1}{2}}\bigcup\mathcal{V}_{\frac{t+1}{2}}\bigcup\mathcal{V}_{\frac{t+3}{2}},\mathcal{E})$. It is well known that this maximal problem is an NP-hard and its complexity is very high since there are ${K/2 \choose (t+1)/2}{K/2 \choose (t+1)/2}+{K/2 \choose (t-1)/2}{K/2 \choose (t+3)/2}+{K/2 \choose (t+3)/2}{K/2 \choose (t-1)/2}$ vertices. We will propose a local maximal matching method to reduce the complexity.

Denote $\mathcal{K}_A=\{a_1$, $a_2$, $\ldots$, $a_{K_A}\}$ and $\mathcal{K}_B=\{b_1$, $b_2$, $\ldots$, $b_{K_B}\}$. We divide sets $\mathcal{V}_{\frac{t-1}{2}}$, $\mathcal{V}_{\frac{t+1}{2}}$ and $\mathcal{V}_{\frac{t+3}{2}}$ into four subsets respectively in the following way:
\begin{small}
\begin{eqnarray}
\label{eq-four-subsets}
\begin{split}
\mathcal{V}_{w;a_1,b_1 }=\{\mathcal{S}\in \mathcal{V}_{w}\ |\  a_1\in \mathcal{S}, b_1\in\mathcal{S} \}, \ \ \
\mathcal{V}_{w;a_1 ,\overline{b}_1}=\{\mathcal{S}\in \mathcal{V}_{w}\ |\ a_1\in \mathcal{S}, b_1\not\in\mathcal{S} \},\\
\mathcal{V}_{w;\overline{a}_1,b_1}=\{\mathcal{S}\in \mathcal{V}_{w}\ |\ a_1\not\in \mathcal{S}, b_1\in\mathcal{S} \},\ \ \
\mathcal{V}_{w;\overline{a}_1 ,\overline{b}_1 }=\{\mathcal{S}\in \mathcal{V}_{w}\ |\ a_1\not\in \mathcal{S}, b_1\not\in\mathcal{S} \},
\end{split}
\end{eqnarray}
\end{small}
where $w=\frac{t-1}{2}$, $\frac{t+1}{2}$, $\frac{t+3}{2}$. It is easy to check that
\begin{small}
\begin{eqnarray}
\label{eq-subsets-cardinality}
\begin{split}
|\mathcal{V}_{w;a_1,b_1}|={K/2 -1\choose w-1}{K/2 -1\choose t-w},\ \ \ \ \ \
|\mathcal{V}_{w;a_1,\overline{b}_1 }|={K/2 -1\choose w-1}{K/2 -1\choose t+1-w},\\[0.2cm]
|\mathcal{V}_{w;\overline{a}_1 ,b_1}|={K/2 -1\choose w}{K/2 -1\choose t-w},\ \ \ \ \ \
|\mathcal{V}_{w;\overline{a}_1 ,\overline{b}_1 }|={K/2 -1\choose w}{K/2 -1\choose t+1-w}.
\end{split}
\end{eqnarray}
\end{small}


Let $\lambda=\frac{M}{N}$.   Given a fixed number $\lambda$, Table \ref{tab-cardinality} can be obtained by \eqref{eq-subsets-cardinality} when $K$ is appropriate large.
\begin{table}
\caption{The cardinality of subsets in \eqref{eq-four-subsets}\label{tab-cardinality}}
\begin{center}\setlength{\arraycolsep}{0.1pt}
\begin{tabular}{|c|c|c|c|}

\hline
Subsets& Cardinality & $|\mathcal{V}_{w;,}|/|\mathcal{V}_{\frac{t+1}{2};\overline{a}_1 ,\overline{b}_1 }|$ &  $|\mathcal{V}_{w;,}|/|\mathcal{V}_{\frac{t+1}{2};\overline{a}_1 ,\overline{b}_1 }|\approx$  \\
\hline

$\mathcal{V}_{\frac{t-1}{2};a_1,b_1}$ & ${K/2 -1\choose (t-3)/2}{K/2 -1\choose (t+1)/2}$& $\frac{(t+1)(t-1)}{(K-t+1)(K-t-1)}$ & $\frac{\lambda^2}{(1-\lambda)^2}$ \\ \hline
$\mathcal{V}_{\frac{t-1}{2};a_1,\overline{b}_1 }$&${K/2 -1\choose (t-3)/2}{K/2 -1\choose (t+3)/2}$& $\frac{(t+1)(t-1)(K-t-3)}{(t+3)(K-t+1)(K-t-1)}$&$\frac{\lambda}{1-\lambda}$\\ \hline
$\mathcal{V}_{\frac{t-1}{2};\overline{a}_1 ,b_1}$&${K/2 -1\choose (t-1)/2}{K/2 -1\choose (t+1)/2}$& $\frac{t+1}{K-t-1}$ & $\frac{\lambda}{1-\lambda}$\\ \hline
$\mathcal{V}_{\frac{t-1}{2};\overline{a}_1 ,\overline{b}_1 }$& ${K/2 -1\choose (t-1)/2}{K/2 -1\choose (t+3)/2}$& $\frac{(t+1)(K-t-3)}{(t+3)(K-t-1)}$ & $1$\\ \hline

$\mathcal{V}_{\frac{t+1}{2};a_1,b_1}$& ${K/2 -1\choose (t-1)/2}{K/2 -1\choose (t-1)/2}$&$\frac{(t+1)(t+1)}{(K-t-1)(K-t-1)}$& $\frac{\lambda^2}{(1-\lambda)^2}$\\ \hline
$\mathcal{V}_{\frac{t+1}{2};a_1,\overline{b}_1 }$&${K/2 -1\choose (t-1)/2}{K/2 -1\choose (t+1)/2}$&$\frac{t+1}{K-t-1}$& $\frac{\lambda}{1-\lambda}$\\ \hline
$\mathcal{V}_{\frac{t+1}{2};\overline{a}_1 ,b_1}$&${K/2 -1\choose (t+1)/2}{K/2 -1\choose (t-1)/2}$&$\frac{t+1}{K-t-1}$&$\frac{\lambda}{1-\lambda}$\\ \hline
$\mathcal{V}_{\frac{t+1}{2};\overline{a}_1 ,\overline{b}_1 }$&${K/2 -1\choose (t+1)/2}{K/2 -1\choose (t+1)/2}$&$1$&$1$\\ \hline

$\mathcal{V}_{\frac{t+3}{2};a_1,b_1}$& ${K/2 -1\choose (t+1)/2}{K/2 -1\choose (t-3)/2}$& $\frac{(t+1)(t-1)}{(K-t+1)(K-t-1)}$ & $\frac{\lambda^2}{(1-\lambda)^2}$\\ \hline
$\mathcal{V}_{\frac{t+3}{2};a_1,\overline{b}_1 }$&${K/2 -1\choose (t+1)/2}{K/2 -1\choose (t-1)/2}$& $\frac{t+1}{K-t-1}$ &$\frac{\lambda}{1-\lambda}$\\ \hline
$\mathcal{V}_{\frac{t+3}{2};\overline{a}_1 ,b_1}$&${K/2 -1\choose (t+3)/2}{K/2 -1\choose (t-3)/2}$& $\frac{(t+1)(t-1)(K-t-3)}{(t+3)(K-t+1)(K-t-1)}$&$\frac{\lambda}{1-\lambda}$\\ \hline
$\mathcal{V}_{\frac{t+3}{2};\overline{a}_1 ,\overline{b}_1 }$ & ${K/2 -1\choose (t+3)/2}{K/2 -1\choose (t-1)/2}$& $\frac{(t+1)(K-t-3)}{(t+3)(K-t-1)}$ & $1$\\ \hline
\end{tabular}
\end{center}
\label{match2}
\end{table}

Now let us consider the sets in \eqref{eq-bi-XY} and their subsets in \eqref{eq-four-subsets}. We can obtain a bipartite graph for any two different subsets. However we only interested in the bipartite graph which has at least one edge. It is not difficult to check that any two elements of a set can not form an effective pair since they have the same number of users requiring from server $A$(and sever $B$). So we only need to consider any two subsets from distinct sets. We take bipartite graph $\mathbf{G}=(\mathcal{V}_{\frac{t+1}{2};\overline{a}_1,\overline{b}_1}, \mathcal{V}_{\frac{t-1}{2};\overline{a}_1,b_1};\mathcal{E})$ as an example.
First let us count the degree of each vertex in $\mathcal{V}_{\frac{t+1}{2};\overline{a}_1,\overline{b}_1}$. Given a vertex $$\mathcal{S}=\{a_{i_1},a_{i_2},\ldots,a_{i_{(t+1)/2}},b_{i'_1},b_{i'_2},\ldots,b_{i'_{(t+1)/2}}\}\in \mathcal{V}_{\frac{t+1}{2};\overline{a}_1 ,\overline{b}_1 },$$
$a_1\not\in\{i_1,\ldots,i_{(t+1)/2},i'_1,\ldots,i'_{(t+1)/2}\}$, it is adjacent to $(t+1)/2$ vertices $$\mathcal{S}_{j}=\mathcal{S}\bigcup\{b_1\}\setminus\{a_{i_j}\}\in\mathcal{V}_{\frac{t-1}{2};\overline{a}_1 ,b_1} $$
Hence $d(\mathcal{V}_{\frac{t+1}{2};\overline{a}_1 ,\overline{b}_1})=(t+1)/2$. Now let us consider the degree of each vertex in $\mathcal{V}_{\frac{t-1}{2};\overline{a}_1 ,b_1}$.
Given a vertex $$\mathcal{S}=\{a_{i_1},a_{i_2},\ldots,a_{i_{(t-1)/2}},b_{i'_1},b_{i'_2},\ldots,b_{i'_{(t+1)/2},b_1}\}\in \mathcal{V}_{\frac{t-1}{2};\overline{a}_1 ,b_1},$$
$a_1\not\in\{i_1,\ldots,i_{(t-1)/2},i'_1,\ldots,i'_{(t+1)/2}\}$, it is adjacent to
$\frac{K}{2} -\frac{t+1}{2} =\frac{K-t-1}{2}$ vertices, i.e., $\mathcal{S}\bigcup\{a\}\setminus\{b_1\}\in \mathcal{V}_{\frac{t+1}{2};\overline{a}_1 ,\overline{b}_1 }$ for each $a\in \mathcal{K}_A\setminus(\{a_1\}\bigcup\mathcal{S})$. That is $d(\mathcal{V}_{\frac{t-1}{2};\overline{a}_1 ,b_1})=\frac{K-t-1}{2}$. Similarly we can compute the degree of each vertex in the bipartite graph $\mathbf{G}=(\mathcal{X},\mathcal{Y},\mathcal{E})$ generated by any two subsets from distinct sets and list them in Tables \ref{each-point-degree} and \ref{each-point-degree1} where $d(\mathcal{X})$ and $d(\mathcal{Y})$ are respectively on the the top and bottom of the diagonal in the entry indexed by $\mathcal{X}$ and $\mathcal{Y}$. It is easy to check that
the elements on the top and bottom of the diagonal in the entry indexed by
$(\mathcal{V}_{\frac{t-1}{2};\overline{a}_1,b_1}, \mathcal{V}_{\frac{t+1}{2};\overline{a}_1,\overline{b}_1})$ are $\frac{t+1}{2}$ and $\frac{K-t-1}{2}$ respectively. By the way the entry is defined by empty when there is no edges in the related bipartite graph.
\begin{table}
\caption{The degrees for each bipartite graph (I)\label{each-point-degree}}
\begin{center}
\begin{tabular}{|c|c|c|c|c|c|c|c|c|}
\hline 
\scalebox{0.48}{\diagbox[dir=SE]{$\mathcal{V}$}{\diagbox[dir=SE,height=3em]{$d(\textstyle \mathcal{V})$}{$d(\textstyle \mathcal{V}_{\frac{t+1}{2}}) $}}{$\mathcal{V}_{\frac{t+1}{2}}$}}& \scalebox{0.45}{$\mathcal{V}_{\frac{t+1}{2},a_1,b_1}$}& \scalebox{0.48}{$\mathcal{V}_{\frac{t+1}{2},a_1,\overline{b}_1 }$ }&  \scalebox{0.48}{$\mathcal{V}_{\frac{t+1}{2},\overline{a}_1 ,b_1}$ }&  \scalebox{0.48}{$\mathcal{V}_{\frac{t+1}{2},\overline{a}_1 ,\overline{b}_1 }$}  \\
\hline
\scalebox{0.48}{ $\mathcal{V}_{\frac{t-1}{2},a_1,b_1}$ }& \scalebox{0.48}{\backslashbox{ $\frac{(K-t+1)(t+1)}{4}$}{$\! \frac{(t-1)(K-t-1)}{4}$ }} &
\scalebox{0.48}{\backslashbox{ $\frac{(K-t+1)(t+3)}{4}$}{$\frac{(t-1)(K-t-3)}{4}$ }} & & \\
\hline
\scalebox{0.48}{ $\mathcal{V}_{\frac{t-1}{2},a_1,\overline{b}_1 }$ }& & \scalebox{0.48}{\backslashbox{$\frac{(K-t+1)(t+3)}{4}$}{ $\frac{(t-1)(K-t-3)}{4}$ }} & & \\
\hline
\scalebox{0.48}{ $\mathcal{V}_{\frac{t-1}{2},\overline{a}_1 ,b_1}$ }&\scalebox{0.48}{\backslashbox{$ \;\qquad \frac{t+1}{2} \;$ }{ $ \; \frac{K-t-1}{2}  \qquad \;$ }} &
\scalebox{0.48}{\backslashbox{$ \qquad \quad 1 \qquad $ }{ $ \qquad 1 \qquad \quad$ }} &
\scalebox{0.48}{\backslashbox{$\frac{(K-t-1)(t+1)}{4}$ }{ $\frac{(K-t-1)(t+1)}{4}$ }} &
\scalebox{0.48}{\backslashbox{$ \; \qquad \frac{K-t-1}{2} \;$ }{ $\; \frac{t+1}{2}  \qquad \;$}} \\
\hline
\scalebox{0.48}{ $\mathcal{V}_{\frac{t-1}{2},\overline{a}_1 ,\overline{b}_1 }$ }& & \scalebox{0.45}{\backslashbox{$ \qquad \frac{t+3}{2} \;\;$ }{ $\;\; \frac{K-t-3}{2} \qquad $ }}  & & \scalebox{0.48}{\backslashbox{$\frac{(K-t-3)(t+3)}{4}$ }{ $\frac{(t+1)(K-t-3)}{4}$ }} \\
\hline
\scalebox{0.48}{ $\mathcal{V}_{\frac{t+3}{2},a_1,b_1}$ }& \scalebox{0.48}{\backslashbox{$\frac{(t+1)(K-t+1)}{4}$ }{ $\frac{(t-1)(K-t-1)}{4}$ }}& &
\scalebox{0.48}{\backslashbox{$ \qquad \frac{K-t+1}{2} \;$ }{ $\; \frac{t-1}{2}  \qquad $ }} & \\
\hline
\scalebox{0.48}{ $\mathcal{V}_{\frac{t+3}{2},a_1,\overline{b}_1 }$ }&\scalebox{0.48}{\backslashbox{$ \qquad \frac{t+1}{2} \;\;$ }{ $\;\; \frac{K-t-1}{2} \qquad $ }} &
\scalebox{0.48}{\backslashbox{$\frac{(t+1)(K-t-1)}{4}$ }{ $\frac{(K-t-1)(t+1)}{4}$ }} &
\scalebox{0.48}{\backslashbox{$ \qquad \quad 1 \qquad$ }{ $\qquad 1 \qquad \quad $ }} &
\scalebox{0.45}{\backslashbox{$ \qquad \frac{K-t-1}{2} \;\; $ }{ $\;\; \frac{t+1}{2}  \qquad $ }} \\
\hline
\scalebox{0.48}{ $\mathcal{V}_{\frac{t+3}{2},\overline{a}_1 ,b_1}$ }& & & \scalebox{0.48}{\backslashbox{ $\frac{(t+3)(K-t+1)}{4}$ }{ $\frac{(K-t-3)(t-1)}{4}$}}& \\
\hline
\scalebox{0.48}{ $\mathcal{V}_{\frac{t+3}{2},\overline{a}_1 ,\overline{b}_1 }$ }& & & \scalebox{0.48}{\backslashbox{ $ \qquad \frac{t+3}{2} \;\; $ }{ $\;\; \frac{K-t-3}{2} \qquad $}} & \scalebox{0.48}{\backslashbox{ $\frac{(t+3)(K-t-1)}{4}$ }{ $\frac{(K-t-3)(t+1)}{4}$}} \\
\hline
\end{tabular}
\end{center}
\end{table}
\begin{table}
\caption{The degrees for each bipartite graph (II)}\label{each-point-degree1}  
\begin{center}  
\begin{tabular}{|c|c|c|c|c|} 
\hline 
\scalebox{0.45}{\diagbox[dir=SE]{$\mathcal{V}_{\frac{t-1}{2}}$}{\diagbox[dir=SE,height=3em]{$\textstyle d(\textstyle \mathcal{V}_{\frac{t-1}{2}})$}{$\textstyle d(\textstyle \mathcal{V}_{\frac{t+3}{2}}) $}}{$\mathcal{V}_{\frac{t+3}{2}}$}}& \scalebox{0.45}{$\mathcal{V}_{\frac{t+3}{2},a_1,b_1}$}& \scalebox{0.45}{$\mathcal{V}_{\frac{t+3}{2},a_1,\overline{b}_1 }$ }&  \scalebox{0.45}{$\mathcal{V}_{\frac{t+3}{2},\overline{a}_1 ,b_1}$}&  \scalebox{0.45}{$\mathcal{V}_{\frac{t+3}{2},\overline{a}_1 ,\overline{b}_1 }$}  \\
\hline
\scalebox{0.45}{ $\mathcal{V}_{\frac{t-1}{2},a_1,b_1}$ }& \scalebox{0.45}{\diagbox{ ${\frac{K-t+1}{2} \choose 2}{\frac{t+1}{2} \choose 2}$}{${\frac{t+1}{2}\choose 2}{\frac{K-t+1}{2} \choose 2}$ }} &
\scalebox{0.45}{\diagbox{ $ \frac{t+1}{2}{\frac{K-t+1}{2} \choose 2}$}{$ \frac{K-t-1}{2}{\frac{t+1}{2}\choose 2}$ }} & & \\
\hline
\scalebox{0.45}{ $\mathcal{V}_{\frac{t-1}{2},a_1,\overline{b}_1 }$ }& & \scalebox{0.45}{\diagbox{${\frac{K-t+1}{2} \choose 2}{\frac{t+3}{2} \choose 2 }$}{ ${\frac{t+1}{2} \choose 2}{\frac{K-t-1}{2} \choose 2}$ }} & & \\
\hline
\scalebox{0.45}{ $\mathcal{V}_{\frac{t-1}{2},\overline{a}_1 ,b_1}$ }&\scalebox{0.45}{\diagbox{$ \frac{K-t-1}{2}{\frac{t+1}{2} \choose 2}$ }{ $\frac{t+1}{2}{\frac{K-t+1}{2} \choose 2}$ }} &
\scalebox{0.45}{\diagbox{$ \frac{(K-t+1)(t+1)}{4}$ }{ $\frac{(t+1)(K-t+1)}{4} $ }} &
\scalebox{0.45}{\diagbox{${\frac{K-t-1}{2} \choose 2}{\frac{t+1}{2} \choose 2}$ }{ ${\frac{t+3}{2} \choose 2}{\frac{K-t+1}{2} \choose 2}$ }} &
\scalebox{0.45}{\diagbox{$ \frac{t+1}{2}{\frac{K-t-1}{2} \choose 2}$ }{ $\frac{K-t-1}{2}{\frac{t+3}{2} \choose 2}$ }} \\
\hline
\scalebox{0.45}{ $\mathcal{V}_{\frac{t-1}{2},\overline{a}_1 ,\overline{b}_1 }$ }& & \scalebox{0.45}{\diagbox{ $\frac{K-t-1}{2}{\frac{t+3}{2} \choose 2}$ }{ $\frac{t+1}{2}{\frac{K-t-1}{2} \choose 2}$}} & &
\scalebox{0.45}{\diagbox{ $ {\frac{K-t-1}{2} \choose 2}{\frac{t+3}{2} \choose 2}$ }{ ${\frac{t+3}{2} \choose 2}{\frac{K-t-1}{2} \choose 2}$}} \\
\hline
\end{tabular}
\end{center}
\end{table}

In order to make the cardinality of a maximal matching as large as possible, we can also use several subsets to generate a bipartite graph.
\begin{example}\rm
\label{ex1}
A bipartite graph $\mathbf{G}=(\mathcal{X},\mathcal{Y}_2;\mathcal{E})$ where
\begin{eqnarray}\label{eq-bi-(3,1)-(5,1)}
\mathcal{X}=\mathcal{V}_{\frac{t+1}{2};\overline{a}_1 ,\overline{b}_1 },\ \ \ \ \ \mathcal{Y}=\mathcal{V}_{\frac{t-1}{2};\overline{a}_1 ,b_1} \bigcup \mathcal{V}_{\frac{t+3}{2};a_1,\overline{b}_1},
\end{eqnarray}
can be obtained. From Table \ref{each-point-degree}, we have $d(\mathcal{V}_{\frac{t+1}{2};\overline{a}_1,\overline{b}_1})=\frac{t+1}{2} + \frac{t+1}{2} = t+1$ and $d(\mathcal{V}_{\frac{t-1}{2};\overline{a}_1,b_1}\bigcup\mathcal{V}_{\frac{t+3}{2};a_1,\overline{b}_1})=\frac{K-t+1}{2}$. Suppose that $\lambda\in (0,\frac{3-\sqrt{5}}{2})$. It is easy to check that $t+1\leq \frac{K-t+1}{2}$ if $\lambda\in (0, \frac{1}{3}-\frac{1}{K}]$, otherwise $t+1\geq\frac{K-t+1}{2}$. From Corollary \ref{co-Hall} there is a saturating matching. So there are $|\mathcal{V}_{\frac{t-1}{2};\overline{a}_1,b_1}\bigcup\mathcal{V}_{\frac{t+3}{2};a_1,\overline{b}_1}|$ or $|\mathcal{V}_{\frac{t+1}{2};\overline{a}_1,\overline{b}_1}|$ vertices in the maximal matching of $\mathbf{G}$ generated by \eqref{eq-bi-(3,1)-(5,1)}. Of course we can also assume that
\begin{eqnarray*}
\mathcal{X}=\mathcal{V}_{\frac{t+1}{2};\overline{a}_1 ,\overline{b}_1 } ,\ \ \mathcal{Y}=\mathcal{V}_{\frac{t-1}{2};\overline{a}_1 ,b_1}
\ \ \ \ \ \hbox{or}\ \ \ \ \
\mathcal{X}=\mathcal{V}_{\frac{t+1}{2};\overline{a}_1 ,\overline{b}_1 },\ \ \mathcal{Y}=\mathcal{V}_{\frac{t+3}{2};a_1,\overline{b}_1}.
\end{eqnarray*}
Similarly we can also show that they have saturating matchings respectively. And it is easy to check that cardinality of the maximal matching by the first assumption is maximal.
\end{example}
With the aid of a computer, we have the following bipartite graphes such that the the number of the unpair of messages is minimal according to the value of $\lambda$.
\subsection{$0< \lambda \leq \frac{3-\sqrt{5}}{2}$}
\label{subsec-1C}
\begin{figure}[htbp]
  \centering
  \includegraphics[width=0.8\textwidth]{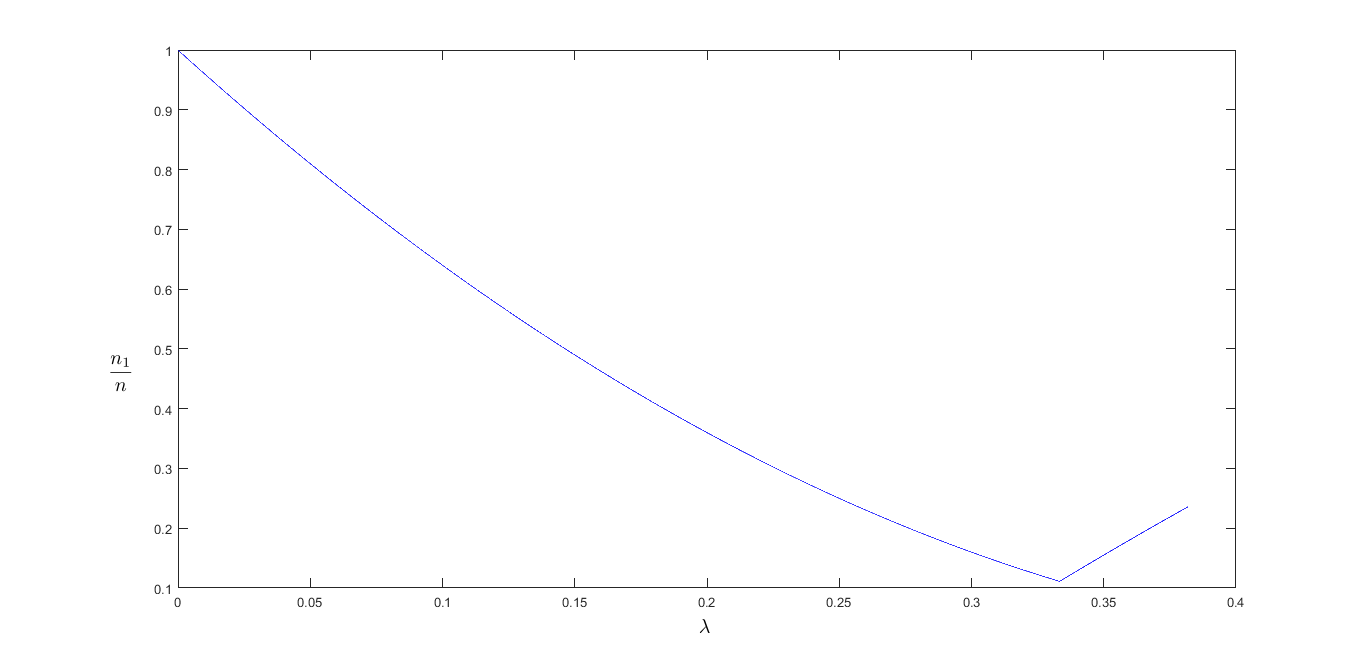}
  \caption{The function $\frac{n_1}{n}$ in \eqref{eq-n1/n} depends on variable $0<\lambda\leq3-\sqrt{5})/2$\label{fig-0-2/5}}
\end{figure}

When $\lambda\leq \frac{3-\sqrt{5}}{2}$, one of the most appropriate method constructing bipartite graphs is 
\begin{eqnarray}
\label{eq-BG1C}
\begin{split}
&\mathbf{G}_1=(\mathcal{V}_{\frac{t+1}{2};\overline{a}_1,\overline{b}_1},
\mathcal{V}_{\frac{t-1}{2};\overline{a}_1,b_1}\bigcup\mathcal{V}_{\frac{t+3}{2};a_1,\overline{b}_1};\mathcal{E}_1)\\[0.2cm]
&\mathbf{G}_2=(\mathcal{V}_{\frac{t+1}{2};\overline{a}_1,b_1},\mathcal{V}_{\frac{t+3}{2};\overline{a}_1,b_1};\mathcal{E}_2)\ \ \ \
\mathbf{G}_3=(\mathcal{V}_{\frac{t+1}{2};a_1,\overline{b}_1},\mathcal{V}_{\frac{t-1}{2};a_1,\overline{b}_1};\mathcal{E}_3)\\[0.2cm]
&\mathbf{G}_4=(\mathcal{V}_{\frac{t-1}{2};\overline{a}_1,\overline{b}_1},
\mathcal{V}_{\frac{t+3}{2};\overline{a}_1,\overline{b}_1};\mathcal{E}_4)\ \ \ \
\mathbf{G}_5=(\mathcal{V}_{\frac{t-1}{2};a_1,b_1},\mathcal{V}_{\frac{t+3}{2};a_1,b_1};\mathcal{E}_5)
\end{split}\end{eqnarray}
From Tables \ref{each-point-degree} and \ref{each-point-degree1}, similar to the discussion in Example \ref{ex1} the following statement holds.
\begin{lemma}\rm
\label{le-BG1C-4C}
Each of the bipartite graphs in \eqref{eq-BG1C} has a saturating matching.
\end{lemma}
From Lemma \ref{le-BG1C-4C} and Table \ref{tab-cardinality}, the number of unpaired messages is
\begin{small}
\begin{eqnarray*}
\label{eq-1C-sum}
\begin{split}
n_1&=\left|\mathcal{V}_{\frac{t+1}{2};a_1,b_1}\right|+\left|\left|\mathcal{V}_{\frac{t+1}{2};\overline{a}_1 ,\overline{b}_1 }\right|-
\left|\mathcal{V}_{\frac{t-1}{2};\overline{a}_1 ,b_1}\bigcup\mathcal{V}_{\frac{t+3}{2};a_1,\overline{b}_1 }|\right|\right|
+\left|\left|\mathcal{V}_{\frac{t+1}{2};\overline{a}_1 ,b_1}\right|-\left| \mathcal{V}_{\frac{t+3}{2};\overline{a}_1 ,b_1}\right|\right|\\[0.2cm]
&+\left|\left|\mathcal{V}_{\frac{t+1}{2};a_1,\overline{b}_1 }\right|-
\left|\mathcal{V}_{\frac{t-1}{2};a_1,\overline{b}_1 }\right|\right|
+\left|\left|\mathcal{V}_{\frac{t-1}{2};\overline{a}_1 ,\overline{b}_1 }\right|-\left| \mathcal{V}_{\frac{t+3}{2};\overline{a}_1 ,,\overline{b}_1 } \right|\right|
+\left|\left|\mathcal{V}_{\frac{t-1}{2};a_1,b_1}\right|-\left| \mathcal{V}_{\frac{t+3}{2};a_1,b_1}\right|\right|\\
&=\left[\left(\textstyle{\frac{t+1}{K-t-1}}\right)^2+\left|1-2{\textstyle\frac{t+1}{K-t-1}}\right|+
2\left|{\textstyle\frac{t+1}{K-t-1}-\frac{(t+1)(t-1)(K-t-3)}{(t+3)(K-t+1)(K-t-1)}}\right|\right]{\textstyle{K/2-1\choose (t+1)/2}^2}\\
&=\left[{\textstyle\frac{|K-3t-3|}{K-t-1}+\frac{(t+1)^2(K-t+1)(t+3)+8K(t+1)(K-t-1)}{(K-t-1)^2(t+3)(K-t+1)}} \right]{\textstyle{K/2-1\choose (t+1)/2}^2}.
\end{split}
\end{eqnarray*}
\end{small}Now let us consider the reduction comparing with the scheme in \cite{LAP}, i.e.,
\begin{eqnarray}
\label{eq-n1/n}
\begin{split}
\frac{n_1}{n}&=\frac{\left[{\textstyle\frac{|K-3t-3|}{K-t-1}+\frac{(t+1)^2(K-t+1)(t+3)+8K(t+1)(K-t-1)}{(K-t-1)^2(t+3)(K-t+1)}} \right]{\textstyle{K/2-1\choose (t+1)/2}^2}}{\left|{K_A\choose (t+1)/2}{K_B\choose (t+1)/2}- {K_A\choose (t-1)/2}{K_B\choose (t+3)/2}-{K_B\choose (t-1)/2}{K_A\choose (t+3)/2}  \right|}\\[0.2cm]
&=\left({\textstyle\frac{|K-3t-3|}{K-t-1}+\frac{(t+1)^2(K-t+1)(t+3)+8K(t+1)(K-t-1)}{(K-t-1)^2(t+3)(K-t+1)}}\right)
\frac{\frac{(K-t-1)^2}{K^2}}{2\frac{(t+1)(K-t-1)}{(K-t+1)(t+3)}-1}\\
&\approx  {\textstyle |1-3\lambda|(1-\lambda)+\lambda^2}.
\end{split}
\end{eqnarray}
The last equation holds when $K$ is appropriate large and $\lambda$ is a fixed number. This implies that the number of unpaired messages left by \eqref{eq-BG1C} is about ${\textstyle \frac{|1-3\lambda|(1-\lambda)+\lambda^2}{3}}$ times smaller than that of obtained by \eqref{eq-bi-XY-L}. Figure \ref{fig-0-2/5} is the function $n_1/n$ depends on variable $\lambda\in (0,\frac{3-\sqrt{5}}{2}]$.
Clearly $n_1/n\approx \frac{1}{9}$ if $\lambda$ towards $\frac{1}{3}$. We can also compute the ratio of unpaired messages as follows.
\begin{eqnarray}
\label{eq delta1-c}
\begin{split}
\Delta_1&=\frac{n_1}{{K \choose t+1}}<\frac{n_1}{\left|\mathcal{V}_{\frac{t+1}{2}}\right|+\left|\mathcal{V}_{\frac{t-1}{2}}\bigcup \mathcal{V}_{\frac{t+3}{2}}\right|}\\
&=\left[{\textstyle\frac{|K-3t-3|}{K-t-1}+\frac{(t+1)^2(K-t+1)(t+3)+8K(t+1)(K-t+1)}{(K-t-1)^2(t+3)(K-t+1)}} \right]\\
&\ \ \ \ \ \ \ \ {\textstyle\frac{{K/2-1\choose (t+1)/2}^2}{2{K/2 \choose (t-1)/2}
{K/2 \choose (t+3)/2}+{K/2 \choose (t+1)/2}{K/2 \choose (t+1)/2}}}\\
&=\left[{\textstyle\frac{|K-3t-3|}{K-t-1}+\frac{(t+1)^2(K-t+1)(t+3)+8K(t+1)(K-t-1)}{(K-t-1)^2(t+3)(K-t+1)}} \right]\frac{\frac{(K-t-1)^2}{K^2}}{2\frac{(t+1)(K-t-1)}{(K-t+1)(t+3)}+1}\\
&\approx {\textstyle \frac{|1-3\lambda|(1-\lambda)+\lambda^2}{3}}
\end{split}
\end{eqnarray}Clearly $\Delta_1$ tends to $\frac{1}{27}$ if $\lambda$ towards $\frac{1}{3}$.

\subsection{$\frac{3-\sqrt{5}}{2}< \lambda \leq \frac{\sqrt{5}-1}{2}$}
\label{subsec-2C}
\begin{figure}[htbp]
  \centering
  \includegraphics[width=0.8\textwidth]{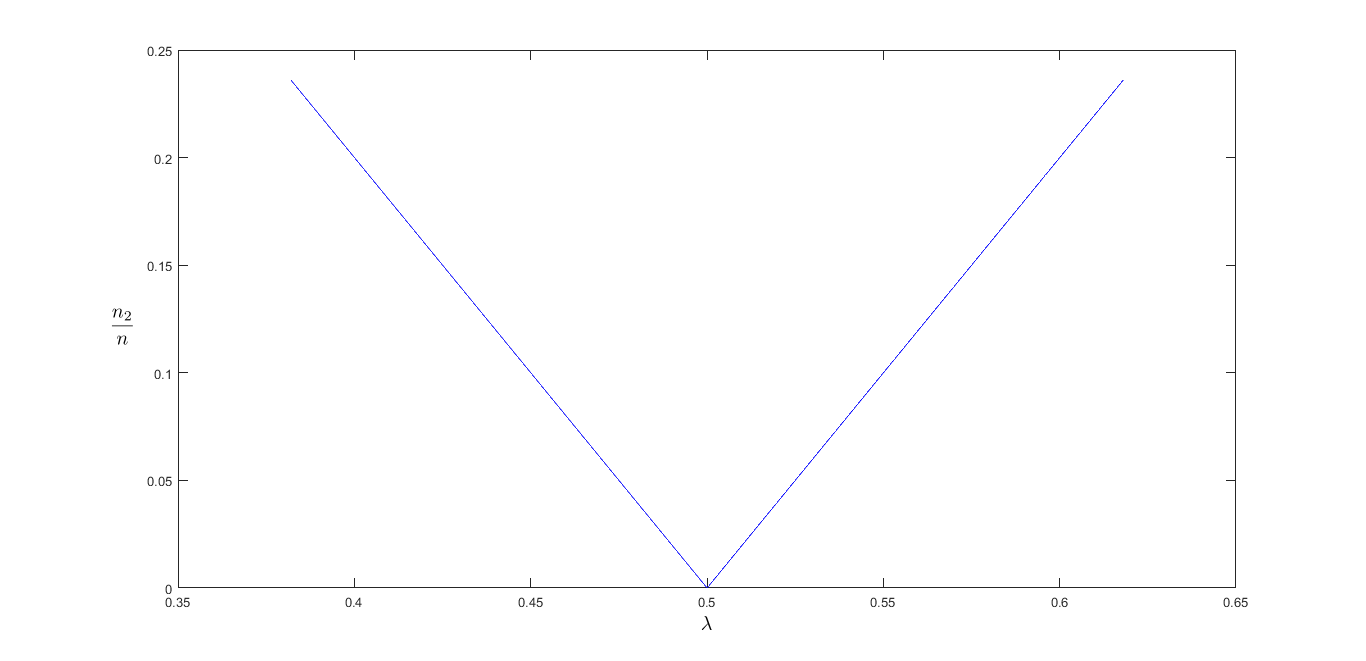}
  \caption{The function $\frac{n_2}{n}$ in \eqref{eq-n2/n} depends on variable $\frac{3-\sqrt{5}}{2}< \lambda \leq \frac{\sqrt{5}-1}{2}$}\label{plan1}
\end{figure}

When $\frac{3-\sqrt{5}}{2}< \lambda \leq \frac{\sqrt{5}-1}{2}$, one of the most appropriate method constructing bipartite graphs is
\begin{eqnarray}
\label{eq-BG2C}
\begin{split}
&\mathbf{G}_1=(\mathcal{V}_{\frac{t+1}{2};a_1,b_1},\mathcal{V}_{\frac{t-1}{2};\overline{a}_1,b_1};\mathcal{E}_1),\ \ \ \ \ \ \mathbf{G}_2=(\mathcal{V}_{\frac{t+1}{2};a_1,\overline{b}_1},\mathcal{V}_{\frac{t-1}{2};a_1,\overline{b}_1};\mathcal{E}_2)\\[0.2cm]
&\mathbf{G}_3=(\mathcal{V}_{\frac{t+1}{2};\overline{a}_1,b_1},\mathcal{V}_{\frac{t+3}{2};\overline{a}_1,b_1};\mathcal{E}_3)\ \ \ \ \ \ \
\mathbf{G}_4=(\mathcal{V}_{\frac{t+1}{2};\overline{a}_1,\overline{b}_1},
\mathcal{V}_{\frac{t+3}{2};a_1,\overline{b}_1};\mathcal{E}_4)\\[0.2cm]
&\mathbf{G}_5=(\mathcal{V}_{\frac{t-1}{2};a_1,b_1},\mathcal{V}_{\frac{t+3}{2};a_1,b_1};\mathcal{E}_5)\ \ \ \ \ \ \
\mathbf{G}_6=(\mathcal{V}_{\frac{t-1}{2};\overline{a}_1,\overline{b}_1},
\mathcal{V}_{\frac{t+3}{2};\overline{a}_1,\overline{b}_1};\mathcal{E}_6)
\end{split}\end{eqnarray}
From Tables \ref{each-point-degree} and \ref{each-point-degree1}, similar to the discussion in Example \ref{ex1} the following result can be obtained.
\begin{lemma}\rm
\label{le-BG2C-4C}
Each of the bipartite graphs in \eqref{eq-BG2C} has a saturating matching.
\end{lemma}
From Lemma \ref{le-BG2C-4C} and Table \ref{tab-cardinality}, the number of unpaired messages is
\begin{small}
\begin{eqnarray*}
n_2&=
\left|\left|\mathcal{V}_{\frac{t+1}{2};a_1,b_1}\right|-\left|\mathcal{V}_{\frac{t-1}{2};\overline{a}_1,b_1}\right|\right|
+\left|\left|\mathcal{V}_{\frac{t+1}{2};a_1,\overline{b}_1}\right|-\left|\mathcal{V}_{\frac{t-1}{2};a_1,\overline{b}_1}\right|\right|
+\left|\left|\mathcal{V}_{\frac{t+1}{2};\overline{a}_1,b_1}\right|-\left|\mathcal{V}_{\frac{t+3}{2};\overline{a}_1,b_1}\right|\right|
\\[0.2cm]
&+\left|\left|\mathcal{V}_{\frac{t+1}{2};\overline{a}_1,\overline{b}_1}\right|-\left|\mathcal{V}_{\frac{t+3}{2};a_1,\overline{b}_1}\right|\right|
+\left|\left|\mathcal{V}_{\frac{t-1}{2};a_1,b_1}\right|-\left|\mathcal{V}_{\frac{t+3}{2};a_1,b_1}\right|\right|+
           \left|\left|\mathcal{V}_{\frac{t-1}{2};\overline{a}_1,\overline{b}_1}\right|-
           \left|\mathcal{V}_{\frac{t+3}{2};\overline{a}_1,\overline{b}_1}\right|\right|.\\[0.2cm]
&=\left({\textstyle\frac{t+1}{K-t-1}\left|\frac{t+1}{K-t-1}-1\right|}+
2\left|{\textstyle\frac{t+1}{K-t-1}-\frac{(t+1)(t-1)(K-t-3)}{(t+3)(K-t+1)(K-t-1)}}\right|+
{\textstyle\left|\frac{t+1}{K-t-1}-1\right|}\right){\textstyle{K/2-1\choose (t+1)/2}^2}\\
&=\left({\textstyle\frac{K}{(K-t-1)^2}|2t+2-K|+\frac{t+1}{K-t-1}\frac{8K}{(t+3)(K-t+1)}}\right){\textstyle{K/2-1\choose (t+1)/2}^2}\\
\end{eqnarray*}
\end{small}Similar to \eqref{eq-n1/n} and \eqref{eq delta1-c}, we have
\begin{eqnarray}
\label{eq-n2/n}
\begin{split}
\frac{n_2}{n}
&=\left({\textstyle\frac{K}{(K-t-1)^2}|2t+2-K|+\frac{t+1}{K-t-1}\frac{8K}{(t+3)(K-t+1)}}\right)
\frac{\frac{(K-t-1)^2}{K^2}}{2\frac{(t+1)(K-t-1)}{(K-t+1)(t+3)}-1}\\
&\approx{\textstyle |2\lambda-1|}
\end{split}
\end{eqnarray}
and the ratio of unpaired messages
\begin{eqnarray}
\label{eq delta2-c}
\begin{split}
\Delta_2&=\frac{n_2}{{K \choose t+1}}<\frac{n_2}{2{K/2 \choose (t-1)/2}
{K/2 \choose (t+3)/2}+{K/2 \choose (t+1)/2}{K/2 \choose (t+1)/2}}\\
&\approx {\textstyle \frac{1}{3}|2\lambda-1|}
\end{split}
\end{eqnarray}
Clearly both $n_2/n$ and $\Delta_2$ tend to $0$ if $\lambda$ towards $1/2$. Figure \ref{plan1} is the function $n_2/n$ depends on variable $\lambda\in (\frac{3-\sqrt{5}}{2},\frac{\sqrt{5}-1}{2}]$.

\subsection{$\frac{\sqrt{5}-1}{2}<\lambda<1$}
\label{subsec-3C}
\begin{figure}[htbp]
  \centering
  \includegraphics[width=0.8\textwidth]{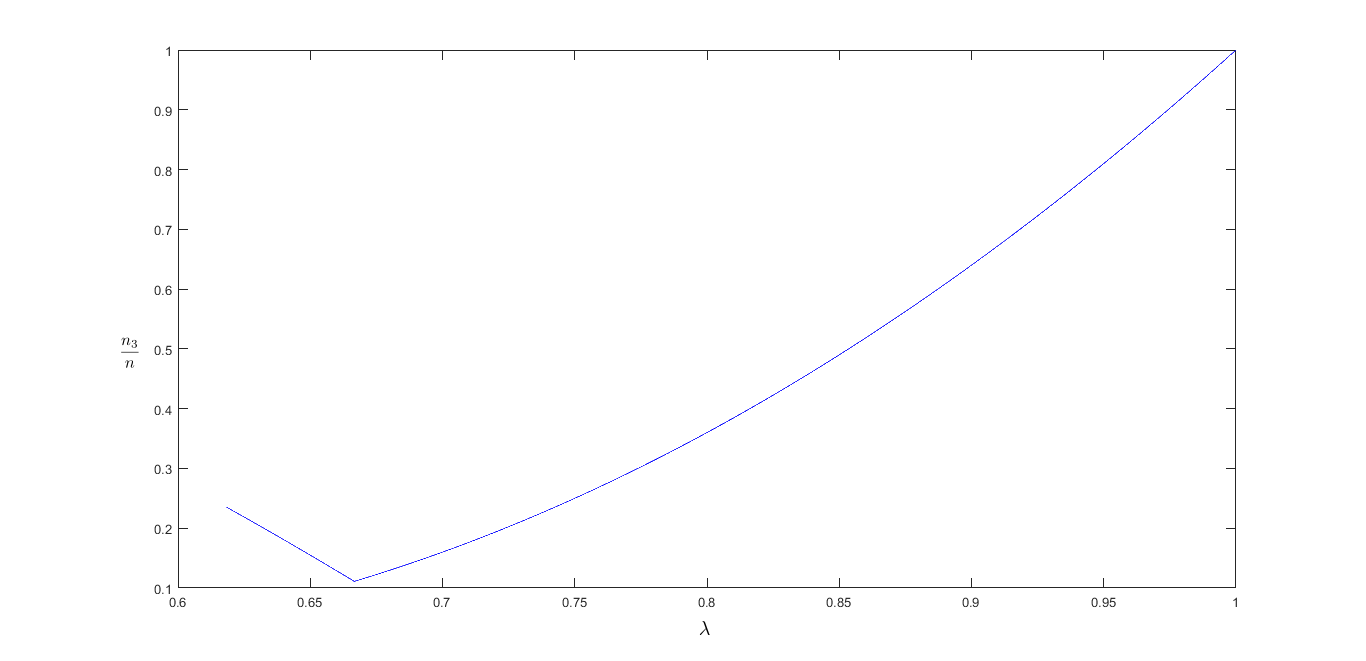}
  \caption{The function $\frac{n_3}{n}$ in \eqref{eq-n3/n} depends on variable $\frac{\sqrt{5}-1}{2}< \lambda < 1$ \label{fig-3/5-1}}
\end{figure}

When $\frac{\sqrt{5}-1}{2}<\lambda<1$, one of the most appropriate method constructing bipartite graphs is
\begin{eqnarray}
\label{eq-BG3C}
\begin{split}
&\mathbf{G}_1=(\mathcal{V}_{\frac{t+1}{2};a_1,b_1},
\mathcal{V}_{\frac{t-1}{2};\overline{a}_1,b_1}\bigcup\mathcal{V}_{\frac{t+3}{2};a_1,\overline{b}_1};\mathcal{E}_1)\\[0.2cm]
&\mathbf{G}_2=(\mathcal{V}_{\frac{t+1}{2};a_1,\overline{b}_1},\mathcal{V}_{\frac{t-1}{2};a_1,\overline{b}_1};\mathcal{E}_2)\ \ \ \
\mathbf{G}_3=(\mathcal{V}_{\frac{t+1}{2};\overline{a}_1,b_1},\mathcal{V}_{\frac{t+3}{2};\overline{a}_1,b_1};\mathcal{E}_3)\\[0.2cm]
&\mathbf{G}_4=(\mathcal{V}_{\frac{t-1}{2};\overline{a}_1,\overline{b}_1},
\mathcal{V}_{\frac{t+3}{2};\overline{a}_1,\overline{b}_1};\mathcal{E}_4)\ \ \ \
\mathbf{G}_5=(\mathcal{V}_{\frac{t-1}{2};a_1,b_1},\mathcal{V}_{\frac{t+3}{2};a_1,b_1};\mathcal{E}_5)
\end{split}\end{eqnarray}
Similar to the discussions in Section \ref{subsec-1C}, the following results can be obtained.
\begin{lemma}\rm
\label{le-BG3C-4C}
Each of the bipartite graphs in \eqref{eq-BG3C} has a saturating matching.
\end{lemma}
From Lemma \ref{le-BG3C-4C} and Table \ref{tab-cardinality}, the number of unpaired messages is
\begin{small}
\begin{eqnarray*}
n_3&=
\left|\mathcal{V}_{\frac{t+1}{2};\overline{a}_1,\overline{b}_1}\right|+\left|\left|\mathcal{V}_{\frac{t+1}{2};a_1,b_1}\right|-
\left|\mathcal{V}_{\frac{t-1}{2};\overline{a}_1,b_1}\bigcup\mathcal{V}_{\frac{t+3}{2};a_1,\overline{b}_1}\right|\right|
+\left|\left|\mathcal{V}_{\frac{t+1}{2};a_1,\overline{b}_1}\right|-\left|\mathcal{V}_{\frac{t-1}{2};a_1,\overline{b}_1}\right|\right|\\[0.2cm]
&+\left|\left|\mathcal{V}_{\frac{t+1}{2};\overline{a}_1,b_1}\right|-
\left|\mathcal{V}_{\frac{t+3}{2};\overline{a}_1,b_1}\right|\right|
+\left|\left|\mathcal{V}_{\frac{t-1}{2};\overline{a}_1 ,\overline{b}_1 }\right|-\left| \mathcal{V}_{\frac{t+3}{2};\overline{a}_1 ,,\overline{b}_1 } \right|\right|
+\left|\left|\mathcal{V}_{\frac{t-1}{2};a_1,b_1}\right|-\left| \mathcal{V}_{\frac{t+3}{2};a_1,b_1}\right|\right|\\
&=\left(1+{\textstyle\frac{t+1}{K-t-1}\left|\frac{t+1}{K-t-1}-2\right|}+
2\left|{\textstyle\frac{t+1}{K-t-1}-\frac{(t+1)(t-1)(K-t-3)}{(t+3)(K-t+1)(K-t-1)}}\right|\right){\textstyle{K/2-1\choose (t+1)/2}^2}\\
&=\left(1+{\textstyle\frac{K}{(K-t-1)^2}|3t+3-2K|+\frac{t+1}{K-t-1}\frac{8K}{(t+3)(K-t+1)}}\right){\textstyle{K/2-1\choose (t+1)/2}^2}\\
\end{eqnarray*}
\end{small}
Similar to \eqref{eq-n1/n} and \eqref{eq delta1-c}, we have
\begin{small}
\begin{eqnarray}
\label{eq-n3/n}
\begin{split}
\frac{n_3}{n}
&=\left(1+{\textstyle\frac{K}{(K-t-1)^2}|3t+3-2K|+\frac{t+1}{K-t-1}\frac{8K}{(t+3)(K-t+1)}}\right)
\frac{\frac{(K-t-1)^2}{K^2}}{2\frac{(t+1)(K-t-1)}{(K-t+1)(t+3)}-1}\\
&\approx{\textstyle (1-\lambda)^2+\lambda\left|3\lambda-2\right|}
\end{split}
\end{eqnarray}
\end{small}and the ratio of unpaired messages left
\begin{small}\begin{eqnarray}
\label{eq delta3-c}
\begin{split}
\Delta_3&=\frac{n_3}{{K \choose t+1}}<\frac{n_3}{2{K/2 \choose (t-1)/2}
{K/2 \choose (t+3)/2}+{K/2 \choose (t+1)/2}{K/2 \choose (t+1)/2}}\\
&=\left[1+{\textstyle\frac{K}{K-t-1}\left|\frac{t+1}{K-t-1}-2\right|+\frac{t+1}{K-t-1}\frac{8K}{(t+3)(K-t+1)}}\right]{\textstyle\frac{{K/2-1\choose (t+1)/2}^2}{2{K/2 \choose (t-1)/2}
{K/2 \choose (t+3)/2}+{K/2 \choose (t+1)/2}{K/2 \choose (t+1)/2}}}\\
&\approx {\textstyle \frac{(1-\lambda)^2+\lambda\left|3\lambda-2\right|}{3}}
\end{split}
\end{eqnarray}\end{small}Clearly $n_3/n$ and $\Delta_3$ tend to $\frac{1}{9}$ and  $\frac{1}{27}$ respectively if $\lambda$ towards $\frac{2}{3}$. Figure \ref{fig-3/5-1} is the function $n_3/n$ depends on variable $\lambda\in (\frac{\sqrt{5}-1}{2},1)$.

By \eqref{eq-R-L-M} the following result can be obtained.
\begin{theorem}\rm
\label{th-impr.B}
Based on MN scheme, when $K_A=K_B$, for the server system in Table \ref{tab-server-system} the rate is
\begin{eqnarray}
\label{eq-rate-2}
R_{T}(K,\textstyle\frac{M}{N})=\left\{\begin{array}{cc}
            \frac{1}{2}R_{MN}(K,\frac{M}{N}) & \hbox{if}\ \frac{KM}{N} \ \hbox{is even}\\
            (\frac{1}{2}+\frac{1}{6}\Delta' )R_{MN}(K,\frac{M}{N}) & \hbox{if}\ \frac{KM}{N} \ \hbox{is odd}
 \end{array}\right.
\end{eqnarray}
where
$$\Delta' \approx
\left \{ \begin{array}{ll}
{\textstyle \frac{|1-3\lambda|(1-\lambda)+\lambda^2}{3}}&\mbox{ If $0< \lambda \leq \frac{3-\sqrt{5}}{2}$}\\
{\textstyle \frac{1}{3}|2\lambda-1|}&\mbox{ If $\frac{3-\sqrt{5}}{2}< \lambda \leq \frac{\sqrt{5}-1}{2}$}\\
{\textstyle \frac{(1-\lambda)^2+\lambda\left|3\lambda-2\right|}{3}}&\mbox{ If $\frac{\sqrt{5}-1}{2}<\lambda<1$}
\end{array}
\right. ,
$$
$\lambda = M/N$,  represents the ratio of unpaired messages.
\end{theorem}

\section{General cases}

In general, we can consider $L$ servers. Basically, we follow the settings of \cite{LAP} where each data segment is stored in a single server, all
caches have the same capacity, and users request a single file. We should indicate that \cite{LAP} discussed the coded caching
system for general multiple servers. However, their methods for all the general cases are based on the three servers scheme
as we discussed
in previous section. Therefore we can just follow their methods to treat the general cases. By this reason, in the
following we just outline
methods for various general cases without going
 into details. The details can be obtained using the method similar to that in \cite{LAP}.

\subsection{Asymmetric requests}

First we consider asymmetric requests for three servers as in Section \ref{sec-Sepcial-idea}. Without loss of generality,
we assume that $K_A > K_B$ and $K_A - K_B = l$. Then we divide ${\mathcal K}_A$ into two parts, one part is of the size $K_B$
and other part is of size $l$. For the first part with ${\mathcal K}_B$, we can use the previous method by pairing the effective
$t+1 -l$ subsets. The remaining request is just like a request for the one server system. Now consider the all possible values
of $l$, we obtain the peak rate as

$$ \sum_{l = 0}^{t+1}{K_A - K_B \choose l}R_T (2K_B, t-l),$$
where $R_T(2K_B, t-l)$ is as in (\ref{eq-rate-2}).

\subsection{More data servers}

Now we consider more data servers. During the off-peak traffic times, for each server the users caches the segments
in a way same
as in the MN scheme. In the requesting time, if a server receives $m$ requests, then it will transmit
$ { K \choose t+1} - {K - m \choose t+1}$ messages. So the normalized peak rate for that server will be
$$\left( { K \choose t+1} - {K - m \choose t+1}\right) \Big/ {K\choose t}.$$

Suppose there are $L$ servers and one parity. Consider two $(t+1)$-subsets of users ${\mathcal S}_1$
and ${\mathcal S}_2$. Using a similar method in the
three servers system, we can let two servers, say $A$ and
$B$, and the parity send one message to the users and other servers send two messages if
two subsets are effective pair.  Similar to previous method, we divide  ${\mathcal S}_1$
into four parts: ${\mathcal Q}_A, {\mathcal Q}_B,{\mathcal Q}'_A, Y$, and ${\mathcal S}_2$
into: ${\mathcal Q}_A, {\mathcal Q}_B,{\mathcal Q}'_B, Y'$, where $Y$ and $Y'$ denote the requests to servers other than
$A$ and $B$.  The other parts are similar to the parts defined in II-B. A little detailed calculation shows the normalized peak
rate in this situation is
$$\frac{(L-1)(K-t)}{L(1+t)}.$$

Now we consider $L$ data servers with two parities $P$ and $Q$. Instead of  using XOR, a higher order field is used to
store data in the parities.  The files stored in parities are shown in Table \ref{tab-L-server-system}, with the assumption
that the servers form an MDS code.

\begin{table}[h]
 \centering
  \caption{Files stored in $P$ and $Q$} \label{tab-L-server-system}
  \normalsize{
\begin{tabular}{|c|c|}
\hline
Server $P$ & Server $Q$ \\ \hline
$A_1+B_1+ \cdots +L_1$      & $A_1+\alpha_BB_1+ \cdots +\alpha_L L_1 $\\
 $A_2+B_2+ \cdots +L_2$      & $A_2+\alpha_BB_2+ \cdots +\alpha_L L_2 $\\

$\vdots$   & $\vdots$        \\
   $A_r+B_r+ \cdots +L_r$      & $A_r+\alpha_BB_r+ \cdots +\alpha_L L_r $\\ \hline
\end{tabular}}
\end{table}

For two $t+1$-subsets of users ${\mathcal S}_1$ and ${\mathcal S}_2$, let $Y = {\mathcal S}_1\cap {\mathcal S}_2$.
Divide ${\mathcal S}_1$ into two parts $Y$ and ${\mathcal Q}'_A$ and divide ${\mathcal S}_2$ into two parts
$Y$ and ${\mathcal Q}'_B$.
Using a similar idea of pairing, server $A$ and $B$ send $m^A_{{\mathcal S}_1}$ and $m^B_{{\mathcal S}_2}$,
respectively. For each
of other servers, sends a message to $Y$ with the content for each user: users requiring files for $B$ received matching
ones so that the desired segments can be decoded using the parity $P$ later, and the remaining users in $Y$ will get
the desired segment corresponding to ${\mathcal S}_1$ when possible, or undesired segment corresponding to
${\mathcal S}_2$. Finally, parity servers $P$ and $Q$ transmit a message to $Y$ with a combination of segments for
each user so that they can recover their required file. Using this method, one can prove that the normalized peak rate
in this situation is:
$$\left( \frac{1}{2} + \frac{L-2}{2L+4}\Delta'\right) \frac{K-t}{t+1}.$$

\subsection{More refined methods}

Finally, we should point out that the rate in Theorem \ref{th-impr.B} can be further improved when $t$ is odd. First let us generalize the notations in \eqref{eq-four-subsets}, i.e., define
\begin{eqnarray}
\label{eq-G-subsets}
\begin{split}
&\mathcal{V}_{w;\overline{a}_1,\ldots,\overline{a}_{h_1-1},a_{h_1},\overline{b}_1,\ldots,\overline{b}_{h_2-1},b_{h_2}}\\
&\ \ \ \ \ \ \ \ =\{\mathcal{S}\in \mathcal{V}_{w}\ |\ a_{h_1}, b_{h_2}\in\mathcal{S}, a_i,b_j\not\in \mathcal{S},i\in[1,h_1), j\in[1,h_2) \}
\end{split}\end{eqnarray}
where $w=\frac{t-1}{2}$, $\frac{t+1}{2}$, $\frac{t+3}{2}$ and $h_1\in[1,\frac{K}{2} -w+1]$, $h_2\in[1,\frac{K}{2} -t+w]$. It is easy to check that
\begin{eqnarray}
\label{eq-G-subsets-cardinality}
|\mathcal{V}_{w;\overline{a}_1,\ldots,\overline{a}_{h_1-1},a_{h_1},\overline{b}_1,\ldots,\overline{b}_{h_2-1},b_{h_2}}|
={K/2 -h_1\choose w-1}{K/2 -h_2 \choose t-w}
\end{eqnarray}
and
$$\mathcal{V}_{w;\overline{a}_1,\ldots,\overline{a}_{h_1-1},a_{h_1},\overline{b}_1,\ldots,\overline{b}_{h_2-1},b_{h_2}}
\bigcap
\mathcal{V}_{w;\overline{a}_1,\ldots,\overline{a}_{h'_1-1},a_{h'_1},\overline{b}_1,\ldots,\overline{b}_{h'_2-1},b_{h'_2}}
=\emptyset$$
for any distinct vectors $(h_1,h_2) \neq (h'_1,h'_2)$. In addition,
\begin{eqnarray}
\label{eq-Com-G-fenchai}
\begin{split}
{K/2 \choose w}{K/2 \choose t+1-w}=\sum^{K/2 -w+1}_{h_1=1}\ \ \sum^{K/2 -t+w}_{h_2=1}
{K/2 -h_1\choose w-1}{K/2 -h_2\choose t-w}
\end{split}
\end{eqnarray}
since it is well know that
\begin{eqnarray*}
{n\choose m}={n-1\choose m-1}+{n-2\choose m-1}+\ldots+{m-1\choose m-1},\ \ \ \ \ \ \ 1\leq m<n
\end{eqnarray*}
always holds. So we have
$$\mathcal{V}_{w}=\bigcup^{K/2 -w+1}_{h_1=1}\ \bigcup^{K/2 -t+w}_{h_2=1}
\mathcal{V}_{w;\overline{a}_1,\ldots,\overline{a}_{h_1-1},a_{h_1},\overline{b}_1,\ldots,\overline{b}_{h_2-1},b_{h_2}}.$$
Then we can also compute the degree of each vertex in the bipartite graph generated by any subsets in \eqref{eq-G-subsets}. Similar to the discussions in Subsections \ref{subsec-1C}, \ref{subsec-2C} and \ref{subsec-3C}, we can further reduce the value of $\Delta'$ in \eqref{eq-rate-2} by sacrificing run-time efficiency on constructing the most appropriate classes of bipartite graphs.
 In fact the sacrificing run-time is very small comparing with that of finding the maximal matching of graph $\mathbf{G} = (\mathcal{V}_{\frac{t-1}{2}}\bigcup\mathcal{V}_{\frac{t+1}{2}}\bigcup\mathcal{V}_{\frac{t+3}{2}};\mathcal{E})$.
\section{Conclusion}
\label{conclusion}
In this paper, we considered the coded caching  scheme for multiple servers setting in \cite{LAP}.  By a refined method for pairing we reduced
the transmission  in the case $\frac{KM}{N}$ is odd for the system with three data servers.
Consequently an obviously smaller rate was obtained. Especially when $K$ is large, $R\approx\frac{1}{2}R_{MN}(K,\frac{M}{N})$ if $\frac{M}{N}$ nears $1/2$, and $R\approx\frac{41}{81}R_{MN}(K,\frac{M}{N})$ if $\frac{M}{N}$ nears $\frac{1}{3}$ or $\frac{2}{3}$.
A brief discussion shows that the improvement can be applied to general multiply servers systems.

In addition, our modification can be generalized to further reduce the rate.
However, with an exhaustive computer search, it will cost more running times to search the bipartite graphs generated by the subsets in \eqref{eq-G-subsets} such that the unpaired messages as small as possible. So it would be of interest if one can found an efficient construction for such bipartite graphs.

\bibliographystyle{model1b-num-names}
\bibliography{<your-bib-database>}

\end{document}